\documentclass[useAMS,usenatbib,usegraphicx]{mn2e}
\usepackage{aas_macros}
\usepackage{color}
\usepackage{array}
\usepackage{amsmath,amssymb}

\begin{document}

  \title[EPTA GWB Limit]{Placing limits on the stochastic gravitational-wave background
  using European Pulsar Timing Array data}
  
  \author[van Haasteren et al.]{
R.~van Haasteren$^1$\footnotemark, Y.~Levin$^{2,1}$, G.~H.~Janssen$^3$, K.~Lazaridis$^4$,
  \\
  \\
    {\LARGE\rm
M.~Kramer$^{4,3}$, B.~W.~Stappers$^{3,5}$, G.~Desvignes$^{6,7,8}$, M.~B.~Purver$^3$,}
  \\
  \\
    {\LARGE\rm
A.~G.~Lyne$^{9}$, R.~D.~Ferdman$^{6,7}$, A.~Jessner$^4$, I.~Cognard$^{6,7}$,}
  \\
  \\
    {\LARGE\rm
    G.~Theureau$^{6,7}$, N.~D'Amico$^{10,11}$, A.~Possenti$^{11}$, M.~Burgay$^{11}$,}
  \\
  \\
    {\LARGE\rm A.~Corongiu$^{11}$, J.~W.~T.~Hessels$^{5,12}$, R.~Smits$^{3,5}$,
    J.~P.~W.~Verbiest$^{4}$}
  \\
  \\
    $^1$Leiden Observatory, Leiden University, P.O. Box 9513, NL-2300 RA
      Leiden, the Netherlands
  \\
    $^2$School of Physics, Monash University, P.O. Box 27, VIC 3800, Australia
  \\
    $^3$University of Manchester, Jodrell Bank Centre for
      Astrophysics, Alan Turing Building, Manchester M13 9PL, UK
  \\
    $^4$Max-Planck-Institut f\"ur Radioastronomie, Auf dem H\"ugel 69,
      53121, Bonn, Germany
  \\
    $^5$Netherlands Institute for Radio Astronomy (ASTRON), Postbus 2, 7990 AA
      Dwingeloo, The Netherlands
  \\
    $^6$LPC2E, Universit\'e d'Orl\'eans - CNRS, 3A Av de la Recherche
      Scientifique, F45071 Orl\'eans Cedex 2, France
  \\
    $^{7}$Station de Radioastronomie de Nan\c{c}ay, Observatoire de Paris,
      CNRS/INSU, F18330 Nan\c cay, France 
  \\
    $^{8}$Department of Astronomy and Radio Astronomy Laboratory, University of
    California, Berkeley, CA 94720, USA
  \\
    $^{9}$University of Manchester, Jodrell Bank Observatory, Macclesfield,
      Cheshire, SK11 9DL, UK
  \\
    $^{10}$Dipartimento di Fisica, Universitá Degli Studi di Cagliari, SP
      Monserrato-Sestu km 0.7, 90042 Monserrato (CA), Italy
  \\
    $^{11}$INAF Osservatorio Astronomico di Cagliari, Loc. Poggio dei Pini, Strada
      54, 09012 Capoterra (CA), Italy
  \\
    $^{12}$Astronomical Institute ``Anton Pannekoek'', University of Amsterdam,
      1098 SJ Amsterdam, The Netherlands}
 
  \date{printed \today}

  \maketitle

  \begin{abstract}
    Direct detection of low-frequency gravitational waves ($10^{-9} - 10^{-8}$ Hz) is
    the main goal of pulsar timing array (PTA) projects. One of the main targets for the
    PTAs is to measure the stochastic background of gravitational waves (GWB) whose
    characteristic strain is expected to approximately follow a power-law of the
    form $h_c(f)=A (f/\hbox{yr}^{-1})^{\alpha}$, where $f$ is the
    gravitational-wave 
    frequency. In this paper we use the current data from the European PTA to
    determine an upper limit on the GWB amplitude $A$ as a function of the
    unknown spectral slope $\alpha$ with a Bayesian algorithm, by modelling the
    GWB as a random Gaussian process.
    For the case $\alpha=-2/3$, which is
    expected if the GWB is produced by supermassive black-hole binaries, we
    obtain a 95\% confidence upper limit on $A$ of $6\times 10^{-15}$, which
    is $1.8$ times lower than the 95\% confidence GWB limit obtained by the
    Parkes PTA in 2006.  Our approach to the data
    analysis incorporates the multi-telescope nature of the European PTA and
    thus can serve as a useful template for future intercontinental PTA
    collaborations.
  \end{abstract}


\footnotetext{Email: haasteren@strw.leidenuniv.nl}

  \section{Introduction}
    The first direct detection of gravitational waves (GWs) would be of great
    importance to astrophysics and fundamental physics: it would  confirm some key predictions of
    general relativity, and  lay the foundation for observational
    gravitational-wave astronomy. Pulsar Timing Arrays (PTAs)  are
    collaborations
    which aim to  detect low-frequency ($10^{-9}$---$10^{-8}$Hz)
    extragalactic gravitational waves directly, by using a set of Galactic
    millisecond pulsars as nearly-perfect Einstein clocks \citep{Foster1990}.
    The basic idea is to exploit the fact that millisecond pulsars create pulse
    trains of exceptional regularity.
    GWs perturb space-time between the pulsars and the Earth, and this creates
    detectable deviations from the strict periodicity in the arrival times of
    the pulses (TOAs) \citep{Estabrook1975, Sazhin1978, Detweiler1979}.

    One of the main astrophysical targets of the PTAs is to measure the stochastic
    background of gravitational waves (GWB). This GWB is expected to be
    generated by a large number of black-hole binaries located at the centres of
    galaxies \citep{Begelman1980, Phinney2001, Jaffe2003, Wyithe2003, Sesana2008},
    by relic gravitational waves \citep{Grishchuk2005}, or, more
    speculatively, by oscillating cosmic-string loops \citep{Damour2005, Olmez2010}.

    Currently, there are three independent PTA groups:\newline
    (i) the Australian-based programme PPTA, the Parkes Pulsar Timing Array, which
    uses data from the Parkes telescope \citep{Hobbs2009, Verbiest2010}, and
    archival Arecibo data.\newline
    (ii) the North-American based programme NANOGrav, North-American Nanohertz
    Observatory for Gravitational waves, which uses both
    the Green Bank Telescope (GBT), and the Arecibo radio telescope
    \citep{Jenet2009}.\newline
    (iii) and the European programme EPTA, European Pulsar Timing Array,
    which uses five different radio telescopes: the Lovell telescope near
    Manchester, United Kingdom, the Westerbork Synthesis Radio Telescope
    (WSRT) in the north of the Netherlands, the Effelsberg Telescope (EFF) near
    Bonn in Germany, the Nan\c cay Radio Telescope (NRT) near Nan\c cay in
    France, and the Sardinia Radio Telescope (SRT) in Sardinia,
    Italy\footnote{The SRT is expected to become operational in 2011
    \citep{Tofani2008}}.\newline
    It is likely that the first detection of GWs by a PTA will occur
    as a result of a joint effort
    of all current PTA projects: an International
    Pulsar Timing Array 
    \citep[IPTA;][]{Hobbs2010}. This will involve the combination of data
    from several different telescopes, each of them with its own specific hardware
    elements and software analysis tools.
    Combining data of different observatories is a challenging task, which
    requires extra care when dealing with the high quality data of modern
    observatories \citep{Janssen2009}.

    In this EPTA paper, we present a methodology on how to combine
    the data from several radio telescopes and use it in an optimal
    way to obtain the information on extragalactic
    gravitational waves. We use the data from three different radio
    telescopes located on the European continent, to place a new upper
    limit on the amplitude of the GWB. As part of our analysis, we
    obtain detailed information about the statistical properties of
    the individual pulse time series.

    The calculation of 
    upper limits on the GWB, based on pulsar
    timing, go as far back as the early 1990's \citep{Stinebring1990, Kaspi1994,
    McHugh1996, Lommen2002}. These analyses have been based on high quality datasets
    for single millisecond pulsars. The most stringent upper limits have been obtained
    recently by \citet{Jenet2006}, who have used PPTA data and archival Arecibo
    data for several millisecond pulsars. Our dataset is different from that
    used by \citet{Jenet2006} since it includes only the pulse times of
    arrival measured by the EPTA telescopes, even though some of the pulsars are
    being timed by multiple PTA groups.
    The Bayesian algorithm we use to obtain an upper limit
    on the GWB is also different from the algorithms used by
    all of the previous studies. Its potential advantages include the use of cross
    correlations between TOAs of different pulsars, and the simultaneous
    constraint on both the amplitude and spectral index of the GWB.

    The outline of the paper is as follows. In Section
    \ref{sec:observations} we give a brief general overview of pulsar
    timing observations. In Section \ref{sec:eptaobs} we detail the
    observations from all of the EPTA telescopes which were used for
    this paper's analysis. 
    We outline the data analysis procedure in Section \ref{sec:bayesian}, after which, in Section \ref{sec:results}, we present the
    upper limits on the amplitude of the GWB, and also the spectral analysis
    of the individual pulsar noises. Finally, in Section
    \ref{sec:implications} we discuss the astrophysical implications of
    our results.


  \section{EPTA data analysis} \label{sec:observations}

    In this section we present a brief overview of the observations,
    instrumentation and data analysis used at the different EPTA
    observatories for transforming a series of measured pulses to a
    TOA. 

\begin{figure*}
	\includegraphics[width=0.9\textwidth]{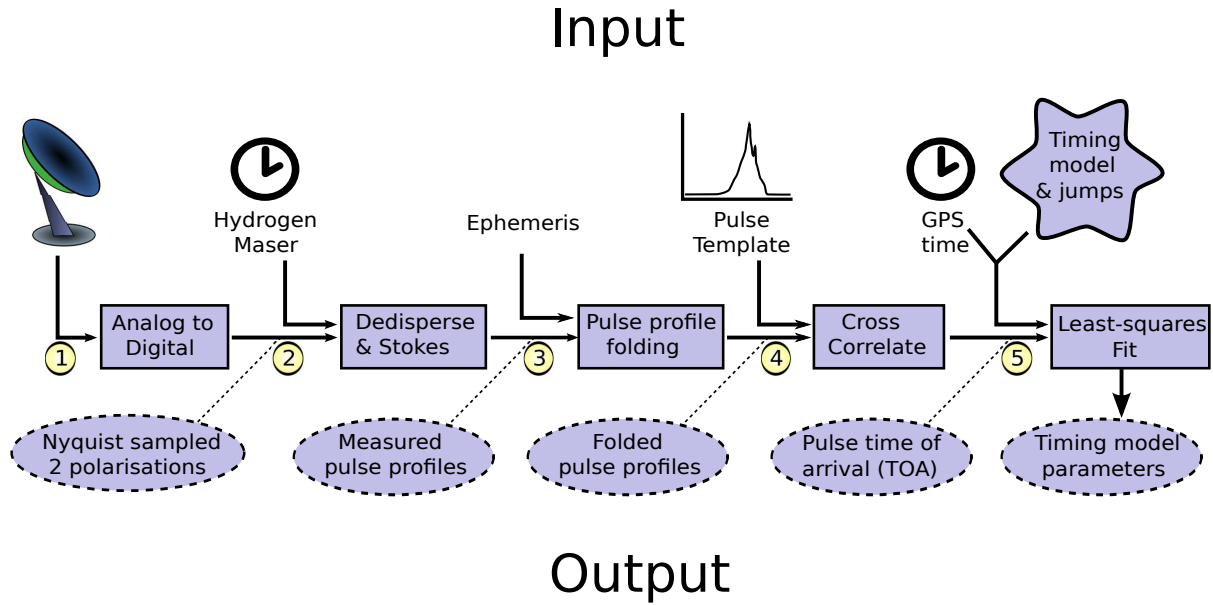}
	\caption{The processing pipeline for pulsar timing, step by step}
	\label{fig:pipeline}
      \end{figure*}

    The complete data reduction process that converts the incoming data stream
    from a radio telescope into one single TOA per observation, called ``the
    pipeline'', is optimised by hand with much care and is observatory specific.
    The process can be described in five general steps, shown in Figure
    \ref{fig:pipeline}:\\
      1) The incoming radio waves are received by the telescope.\\
      2) The signal is converted from analog to digital, at a Nyquist sampled
      rate.\\
      3) Data is (coherently) de-dispersed and, if possible, Stokes parameters
      are formed.\\
      4) The de-dispersed timeseries are folded at the pulsar period, resulting in averaged pulse
      profiles. Typically a timespan containing several $10^5$ pulses is used
      for each TOA.\\
      5) A cross-correlation with a template pulse profile yields a TOA and
      associated uncertainty \citep{Taylor1992}.

   Individual pulse amplitudes and pulse shapes are highly irregular,
   and pulse phases vary significantly from pulse to pulse
   \citep{Cordes2010}.  Therefore careful averaging {\it (folding)}
   has to be performed
   to obtain a single TOA.  Furthermore, the interstellar medium
   (ISM) results in significant delays of the arrival time of the
   pulses over the receiver bandwidth. As a large bandwidth is
   required to reliably detect a pulse, accounting for the ISM
   is key for precision timing.

   Differences in templates used, e.g. the use of integrated
   profiles versus
   analytic templates, all based on single--observatory data, and the
   difference in definition of the reference point in a template will
   result in offsets between data sets generated by different
   observatories. All extra offsets in our data
   will lead to information loss of other signals like the GWB.
   Therefore, using a common template for each pulsar at all
   observatories is desirable, and will be implemented in the near
   future.

   The realisation of the five steps and therefore their output
   (the resulting TOA) might differ among observatories. Understanding
   and accounting for those differences is essential for the correct
   analysis and optimal combining of the EPTA data. A more detailed
   study on this subject is in preparation (Janssen et al. 2011).

    The cross-correlation between the folded profile and the
    template yields an uncertainty of the TOA
    \citep{Taylor1992}. One would like this uncertainty to be solely
    due to the radiometer noise, i.e. the noise intrinsic to the
    measurement, but in practice the errors sometimes appear to have
    been systematically over- or underestimated.  It is a common
    practice, which we follow here, to allow for an extra parameter
    to multiply these uncertainties for each
    pulsar-observatory-backend combination \citep{HobbsTempo}.
    This extra
    multiplicative factor allows the TOA uncertainties to
    statistically account for the TOA scatter: the deviations of the
    strict periodicity of the pulses. This is clearly
    unsatisfactory, and in future timing experiments the origin of
    the predicted and measured TOA scatter will have to be
    thoroughly investigated.

      \begin{table*}
	\centering
	\begin{tabular}{ r  l l l }
	  \hline
           Telescope & WSRT & NRT & EFF \\
	  \hline
	  Equivalent dish size (m) & $93.5$  & $94.4$ & $100$ \\
	  Centre observing frequencies (MHz) & 1380 & 1398,\,2048 & 1400 \\
          Observing bandwidth (MHz) & 80 & 64/128 & 28-112 \\
	  Obs. time per month per pulsar & $1$x$30$ min & $4$-$6$x$60$min & $1$x$30$ min\\
          Pulsar backend & PuMaI & BON & EBPP \\
	  Dedispersion & incoherent & coherent & coherent \\
	  Used templates & integrated profiles & integrated profiles & analytic \\
	  \hline
	\end{tabular}
	\caption{Details of the different EPTA observatories relevant for this
	  work. The NRT observing bandwidth has doubled to 128 MHz in July
	  2009.}
	\label{tab:observatories}
      \end{table*}

    \section{EPTA observations} \label{sec:eptaobs}


  \subsection{Overview of the observatories}

   We have used pulsar timing observations of five radio pulsars, observed with
   three of the EPTA telescopes,
   to
   set a limit on the GWB.  See Table\,\ref{tab:observatories},
   Fig.\,\ref{fig:allresiduals} and the Appendix for an overview of
   the data sets used and the properties of each telescope. Each
   pulsar was observed on average once every month for 30 minutes at
   each telescope. Although additional observing frequencies are
   commonly used at WSRT and EFF, their respective 1380 and 1400\,MHz
   observing bands have the best sensitivity and result in the highest precision
   TOAs. Therefore we have only used observations taken at those
   frequencies at WSRT and EFF for the analysis presented in this
   paper. The data were either coherently de-dispersed (NRT and EFF)
   or incoherently de-dispersed (WSRT).
   The observations
   were folded and cross-correlated with an analytic template (EFF), or a high
   S/N, observatory specific, template (WSRT \& NRT), to calculate one time-of-arrival (TOA) per observation.
   See
   e.g. \citet{Lazaridis2009} for a more
   complete description of the observing procedures and data analysis
   at the different observatories.

   As discussed, any change to the pipeline or to the input of the
   pipeline can result in a difference in the calculated TOAs. We
   emphasise that it is essential to correctly identify these systematic effects
   and include them in the modelling of the TOAs.
   In our analysis, we
   have done this by introducing jumps between TOAs of the same pulsar
   anywhere the pipeline differs in some way.

   Once the complete set of data for each pulsar is obtained, and
   corrected for global drifts by comparing to UTC, it is fit
   with the timing model. The timing model is a multi-parameter fit that
   represents our best knowledge of the many deterministic processes that
   influence the values of the TOAs. The timing residuals are then
   produced by subtracting the timing model, which is subsequently optimised by
   minimising these residuals through a least--squares fit.
   This was done using the pulsar timing package
   {\sc Tempo2} \citep{Hobbs2006}.

      \subsection{Selection of data sets} \label{sec:usedpulsars}
        The European observatories have been timing
        millisecond pulsars for many years, and potentially all
        of that data could be used in the calculation of an upper
        limit on the GWB. However, like \citet{Jenet2006} we choose to use only
	the data from the pulsars which perform best as ideal clocks, e.g. those
	with the highest precision TOAs and the most straight-forward noise
	characteristics.

	TOA precision is not the only factor that determines the sensitivity to
	the GWB; other factors like the total timing baseline and the number of
	observations (i.e. TOAs) affect this sensitivity as well. A great
	advantage of the EPTA data is that several pulsars have been monitored
	for a relatively long time: over $10$ years.
        To determine which timing residuals (i.e. pulsar-observatory
        combinations) are most useful for GWB detection, we analyse each dataset
	separately. By doing this we can determine the sensitivity to
	the GWB of a set of TOAs: the lower the $3$-$\sigma$ upper limit
	$h_c^{\text{max}}(1\text{yr})$ we get using only a particular set of TOAs, the more
	sensitive that set of TOAs is to the GWB.

\begin{figure}
	  \includegraphics[width=0.5\textwidth]{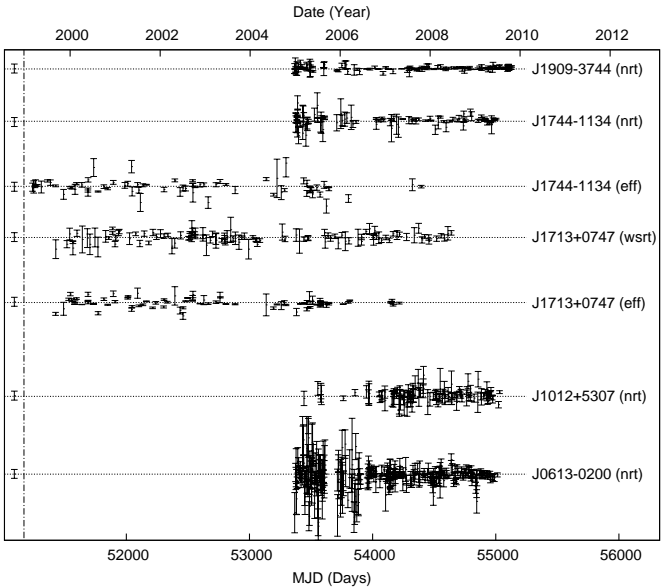}
	  \caption{The timing residuals of all the pulsars used in the GWB limit
	    calculation. The time in MJD is shown on the x-axis. On the left of
	    the dash-dotted line we have placed a sample residual with an
	    uncertainty of $1\,\mu$s. 
          }
	  \label{fig:allresiduals}
	\end{figure}

	The timing residuals of the selected pulsars are shown in Figure
	\ref{fig:allresiduals}. These five pulsars significantly outperform the
	other pulsars being timed by the EPTA in terms of how well they can limit the GWB amplitude:
	these five pulsars can each individually limit the GWB well below
	$h_c(1\text{yr})=10^{-13}$ for
	$\alpha=-2/3$, whereas other current EPTA datasets typically perform
	worse by a factor of several. Since there is such a difference between
	this set of five pulsars, and the other pulsars that have been observed by the EPTA, we do not expect to gain any significant sensitivity by
	including more pulsars that cannot meet this constraint.
	We therefore choose $h_c^{\text{max}}(1\text{yr})\leq 10^{-13}$ with
	$\alpha=-2/3$ as a constraint for including a dataset in our
	calculation.

	In addition to this constraint, we also demand that datasets that
	just barely satisfy $h_c^{\text{max}}(1\text{yr})\leq 10^{-13}$ do not show
	prominent low-frequency (``red'') timing noise.
	Our criterion for
	presence of the latter is a peak in the posterior
	distribution which is inconsistent with zero amplitude for
	$\alpha \leq 0$.

\section{Data analysis} \label{sec:bayesian}
    The analysis presented in this paper broadly follows the procedure introduced in
    \citet[vHLML]{vanhaasteren2009}. The vHLML Bayesian algorithm relies on creating the
    parametrised models of the timing residuals, and forming a probability
    distribution function (PDF) as a function of the model parameters. All known
   systematic contributions of known functional form should be included in the model. In
    the examples used by vHLML  the model for the systematic errors included only
    the quadratic contribution to the TOAs from pulsar spindowns. The
    multi-telescope nature of the EPTA requires  more complete
    models for timing residuals than the one used in vHLML. In this section we
    show how to build and implement these models in practice. 

    We first briefly review the algorithm of vHLML in Section
    \ref{sec:bayesianoverview} and \ref{sec:bayesianresults}. We then present
    the extended model
    we use for the analysis of the TOAs in Section
    \ref{sec:bayesianmodels}, after which we show how we handle TOAs coming from
    different observatories in Section \ref{sec:combining}.

    \subsection{Brief review of the vHLML algorithm} \label{sec:bayesianoverview}
      The set of TOAs from all pulsars forms the basic input used in the Bayesian
      data analysis. Many processes influence the  measured TOA values; in
      this work we discriminate between deterministic processes, like quadratic
      spindown, and stochastic
      processes, like timing noise:
      \begin{equation}
	\label{eq:toaprocesses}
	t_{(ai)}^{\text{obs}} = t_{(ai)}^{\text{det}} + \delta
	t_{(ai)}^{\text{stoch}},
      \end{equation}
      where $t_{(ai)}^{\text{obs}}$ represents the $i$-th TOA of pulsar
      $a$, $t_{(ai)}^{\text{det}}$ is the corresponding contribution to the TOA
      solely due to deterministic processes, and $\delta
      t_{(ai)}^{\text{stoch}}$ is the contribution due to stochastic processes.
      
      The effects of deterministic processes are  described by the set of
      model parameters $\vec{\eta}$: $t^{\text{det}}_{(ai)} =
      t^{\text{det}}_{(ai)}(\vec{\eta})$.  As is done in vHLML, we assume that
      the stochastic processes are Gaussian, though their spectra are not
      necessarily white.  In such a model, the stochastic processes can be
      represented by the correlation matrix
      \begin{eqnarray}
	\label{eq:coherenceensemble}
	\langle\delta t_{(ai)}^{\text{stoch}}\delta
	t_{(bj)}^{\text{stoch}}\rangle = C_{(ai)(bj)} = C_{(ai)(bj)}(\vec{\xi}),
      \end{eqnarray}
      where $\vec{\xi}$ are the model parameters. 

      The key distribution used in a Bayesian analysis is the likelihood function, the
      probability distribution of the data for a given model and its parameters.
      As described in vHLML, for PTAs the likelihood takes the following form:
      \begin{eqnarray} \label{eq:gaussian}
	L\left(\vec{\theta}\right) &=& P\left(\vec{\delta t} \mid \vec{\theta}\right) = 
	  \frac{1}{\sqrt{\left(2\pi\right)^{n} \det C}} \\
	& &\exp\left[-\frac{1}{2}\sum_{{(ai)(bj)}}
	  (\vec{t}^{\text{obs}}_{(ai)}-\vec{t}^{\text{fit}}_{(ai)})
	  C^{-1}_{(ai)(bj)}
	  (\vec{t}^{\text{obs}}_{(bj)}-\vec{t}^{\text{fit}}_{(bj)})\right],\nonumber
      \end{eqnarray}
      where $\vec{\theta} = (\vec{\eta}, \vec{\xi})$, and $\vec{\delta t}$ is
      the difference between the observed TOAs, and the fitted TOAs. A Bayesian
      analysis assigns prior distributions $P_{0}(\vec{\theta})$
      to the model parameters, and explores the parameter space of the
      posterior distribution (short-handed simply as {\it the posterior}):
      $P(\vec{\theta}\mid \vec{\delta t}) =
      L(\vec{\theta})P_{0}(\vec{\theta})$.

    \subsection{Obtaining a marginalised posterior distribution} \label{sec:bayesianresults}
      The posterior $P(\vec{\theta}\mid \vec{\delta t})$ contains  information
      about all model parameters. We need to express the posterior as a
      function of only those parameters that represent the GWB. This process is
      called marginalisation, and consists of integrating over all other
      parameters.  The resulting marginalised posterior is the posterior
      probability density of the GWB parameters. 

      Marginalisation of a posterior in a high-dimensional parameter space is
      non-trivial, and a direct numerical integration is prohibitively
      computationally expensive. As in vHLML, we employ a mix of analytic
      integration and  Markov Chain Monte Carlo (MCMC) methods to accomplish
      this. The marginalisation remains the computational bottleneck for the
      method's effectiveness, as the computational time scales with $n^3$, with
      $n$ the total number of TOAs to be analysed.

      A computational shortcut can be used by analytically marginalising over
      the parameters of the timing model. As shown in vHLML, this is possible
      provided that the parameters represent signals of known functional form.
      This condition is equivalent to the requirement that the timing residuals
      generated by the timing model are linear with respect to its parameters:
      $\delta t = d (\alpha - \hat{\alpha})$, where $\delta t$ is the timing
      residual, $d$ is a proportionality constant, $\hat{\alpha}$ is the best
      fit value for the model parameter, and $\alpha$ is the model parameter.
      While this is always true for quadratic spindown as considered explicitly
      in vHLML, it is generally not true for other timing model parameters.
      However, when the deviations of the timing model parameters from their
      best-fit values are small, it is a good approximation that the residuals
      generated by the timing model are linear with respect to the deviations
      from their best-fit values: $\delta t \approx d (\alpha - \hat{\alpha})$.

      Analytically marginalising over the timing model is therefore possible,
      and by doing so the number of parameters that must be integrated over
      numerically by the use of MCMC is reduced greatly. Dependent on the
      model we use to describe the statistics of the timing residuals, the
      number of parameters left to explore is then just several per
      pulsar/backend combination.
      The results of the analysis can be presented as a marginalised posterior
      as a function of any parameter in the model, provided that this parameter
      was present in the MCMC run.

    \subsection{Used model for the TOAs} \label{sec:bayesianmodels}
      We divide the actual parameterisation in 3 parts:\newline
      a) The deterministic timing model.\newline
      b) The gravitational-wave background.\newline
      c) Other stochastic processes (e.g.,timing noise).\newline
      In this section we discuss how we have taken these into account in our
      data analysis.

      As a first step, the TOAs are processed using the  software package {\sc
      Tempo2},  in order  to determine the best-fit
      timing model. This procedure consists of the
      following steps:\newline
      1. {\sc Tempo2} requires an initial guess $\alpha_{0i}$ for the timing model
      parameters $\alpha_i$ in order to find timing residuals (pre-fit timing
      residuals).\newline
      2. It then constructs an approximation to the timing model, in
      which the timing residuals depend linearly on
      $\alpha_i - \alpha_{0i}$.\newline
      3. It finds the best-fit $\alpha_i$ within this linear
      approximation, and uses those values to update the timing residuals using
      the full non linear timing model (post-fit timing residuals).\newline
      4. The newly obtained parameters and corresponding timing residuals are
      then judged by the person performing the model fitting, and if determined necessary the newly obtained
      parameters can act as the initial guess for a new fitting iteration. {\sc
      Tempo2} also allows adjustment and fitting of $\alpha_i$ one by
      one.\newline
      Finding the timing solution with {\sc Tempo2} is not fully algorithmic, but
      typically requires someone experienced with pulsar timing analysis, who
      approaches the TOAs fitting in several different ways, which ensures that
      phase coherence is maintained and that the relevant deterministic model
      parameters are included properly. Though this strategy works well in practice, we
      should remain conscious of the possibility that different solutions might
      be obtained by different observers, who may also choose to include
      additional model parameters.
      \footnote{Qualitatively, experienced observers are rightfully so very
      confident in their timing solutions. Quantitatively however, the only
      statistical tool currently available for observers to check whether the
      timing solution is reasonable is the reduced $\chi^2$ statistic. But since the
      error bars obtained with the cross-correlation technique cannot be fully
      trusted, the same holds for the $\chi^2$ statistic.} In the appendix we
      present the timing solutions we found for the analysed pulsars. These are
      the values we used as our initial guess, $\alpha_{0i}$.
      Note
      that these $\alpha_{0i}$ and their uncertainties, although created with
      {\sc Tempo2} using the same datasets that we base our upper limit on,
      do not include our model for
      the red noise.
      The values and uncertainties we list in the appendix
      therefore do not represent our best estimates if we were to take into
      account the red timing noise. Although calculating these best estimates of
      $\alpha_i$ is reasonably straightforward, these estimates are not
      accessible in our MCMC because we have marginalised over these parameters
      analytically. The calculated upper limit on the GWB,
      however, does include all these effects, and therefore
      automatically incorporates the removal of power from the low-frequency GW
      signal by fitting for the timing model parameters and jumps.

      In the above mentioned step 2 where the timing model is linearised, we
      have made an important simplification that we now describe in more detail. Since
      we take into account, and marginalise over, all timing model parameters in
      our algorithm, we are effectively working with the TOAs instead of
      just the timing residuals. However, the timing model has been linearised
      by {\sc Tempo2} with respect to $\alpha_i - \alpha_{0i}$. This implies
      that we need to be sufficiently close to $\alpha_{0i}$ in the parameter
      space for this approximation to be valid, which means that the
      timing residuals derived with {\sc Tempo2} need to be approved by the
      person fitting the data, before using these as inputs in the Bayesian algorithm.

      The stochastic component contributing to the TOAs is characterised as
      follows. Firstly, general relativity describes how the timing residuals of
      a pair of pulsars are correlated due to gravitational waves:
      \begin{equation}
	\zeta_{ab}={3 \over 2}\frac{1-\cos\theta_{ab}}{2}
	  \ln\left(\frac{1-\cos\theta_{ab}}{2}\right)
	  -\frac{1}{4}\frac{1-\cos\theta_{ab}}{2}+\frac{1}{2}
	  +\frac{1}{2}\delta_{ab},
	\label{eq:zetaab}
      \end{equation}
      where  $\theta_{ab}$ is the angle between pulsar $a$ and pulsar $b$
      \citep{Hellings1983}.
      The GWB spectrum is parametrised as a power-law of the
      form \citep{Maggiore2000, Phinney2001, Jaffe2003, Wyithe2003, Sesana2008}:
      \begin{eqnarray} \label{eq:charstrain}
	h_c&=&A\left(f\over \hbox{yr}^{-1}\right)^{\alpha},
      \end{eqnarray}
      were $h_c$ is the characteristic strain as used in \citet{Jenet2006},
      $A$ is the amplitude of the signal, and $\alpha$ is the spectral index.
      This then results in a correlation matrix for the GWB (vHLML):
      \begin{eqnarray}
	C^{\rm GW}_{(ai)(bj)} &=&
	  \frac{-A^2\zeta_{ab}}{\left(2\pi\right)^{2}f_{L}^{2-2\alpha}}
	  \left\{\Gamma(-2+2\alpha)\cos\left(\pi\alpha\right)
	  \left(f_{L}\tau\right)^{2-2\alpha} \right.\nonumber\\
	& &\left.\sum_{n=0}^{\infty}\left(-1\right)^{n}
	  \frac{\left(f_{L}\tau\right)^{2n}}{(2n)!
	  \left(2n+2\alpha-2\right)}\right\} ,
	\label{eq:CGW}
      \end{eqnarray}
      where, as in vHLML, $\tau = 2\pi |t_i - t_j|$, and $f_L$ is a cut-off frequency,
      set much lower than the lowest GW frequency we are sensitive to.

      Secondly, the stochastic timing noise for each individual pulsar is split
      into three components:\newline
      1) Individual errors of TOA determination from the cross-correlation,
      represented by the TOA error bars. An extra free parameter, called the
      EFAC value, is commonly introduced by pulsar observers in order to account
      for possible mis-calibration of the radiometer noise \citep{HobbsTempo}; this
      parameter is a multiplier for all of the TOA error bars for a given
      pulsar.\newline
      2) An extra white noise component, independent of the error bars. This
      basically acts as extra non-time--dependent noise, and the parameter is
      often called an EQUAD parameter.\newline
      3) Red noise, consisting of a power-law spectrum in the timing residuals.
      This component allows for structure in the timing residuals.\newline
      All three timing noise components are uncorrelated between the pulsars.

      The resulting correlation matrices from components 1, 2, and 3, as derived
      in vHLML, are given by:
      \begin{eqnarray}
	C^{\rm err}_{(ai)(bj)} &=& E^{2}_{a}\Delta t_{(ai)}^2\delta_{ab}\delta_{ij}
	\nonumber \\
	C^{\rm WN}_{(ai)(bj)} &=& N^{2}_{a}\delta_{ab}\delta_{ij}
	\nonumber \\
	C^{\rm RN}_{(ai)(bj)} &=&
	  \frac{-R^2_{a}\delta_{ab}}{\left(2\pi\right)^{2}f_{L}^{2-2\alpha_{a}}}
	  \left\{\Gamma(-2+2\alpha_{a})\cos\left(\pi\alpha_{a}\right)
	  \left(f_{L}\tau\right)^{2-2\alpha_{a}} \right. \nonumber \\
	& &\left.\sum_{n=0}^{\infty}\left(-1\right)^{n}
	  \frac{\left(f_{L}\tau\right)^{2n}}{(2n)!
	  \left(2n+2\alpha-2\right)}\right\},
	\label{eq:noisematrix}
      \end{eqnarray}
      where $C^{\rm err}_{(ai)(bj)}$, $C^{\rm WN}_{(ai)(bj)}$, and $C^{\rm RN}_{(ai)(bj)}$
      are
      the correlation matrices corresponding to the error bars, the extra white
      noise, and the red noise respectively, with $a$ and $b$ denoting the
      pulsar number, $i$ and $j$ denote the observation number, $\Delta t$ is
      the TOA uncertainty (the error bar) as calculated in the pipeline, $E_{a}$
      is the scaling parameter of the error bars for the $a$'th pulsar (the EFAC factor),  $N_{a}$ is the 
      amplitude of the white noise, $R_{a}$ is the amplitude of the red timing noise, $\alpha_a$ is the spectral index of the red noise spectrum of
      pulsar $a$, and $\tau$ is the time difference between
      two observations.

    \subsection{Combining datasets} \label{sec:combining}
      The previous section gives a complete description of the model we use to
      analyse the TOAs of a single pulsar, observed with one telescope. That
      model does not yet account for the use of different observatories. In this
      section we explain what we do to accomplish this.

      As discussed in Section \ref{sec:observations}, the reduced data products
      are (sometimes subtlety) influenced by many different components of the
      reduction process. In order to
      account for slight offsets between TOAs, 
      introduced by using slightly different reduction procedures on individual
      datasets, a calibration term needs to be introduced when
      merging TOAs from different observing systems. This extra calibration term
      takes the form of a ``jump'', an arbitrary phase offset between datasets,
      which is fit for simultaneously with other timing model parameters.
      We
      use the term dataset for any series of TOAs that can be analysed without a
      jump. In practice this is any series of TOAs, of the same pulsar,
      observed with the same hardware elements, and processed with the same
      algorithms, at the same observing frequency. Here we combine $7$ such
      datasets (those shown in Figure \ref{fig:allresiduals}).

      Jumps have been used routinely when
      combining data of different observatories and/or data recorders
      \citep[e.g.,][]{Janssen2009}.
      This allows us to find a single solution for the timing model of a pulsar
      timed by multiple observatories. However, the TOAs produced by pipelines
      at different observatories may have different statistical properties.
      In order to account for this, we allow the stochastic
      contributions in our model discussed in Section \ref{sec:bayesianmodels} to
      vary between datasets:\newline
      1) One timing model per pulsar (taken directly from {\sc Tempo2})\newline
      2) Jumps between different datasets\newline
      3) A scaling factor for the error bars (EFAC) for each dataset\newline
      4) An extra white noise component (EQUAD) for each dataset\newline
      5) Power law red noise for each dataset\newline
      A major advantage of this approach is that  it allows one to detect
      statistical differences between observatories that may be introduced by
      different algorithms/components, and then use this feedback to iteratively
      improve our datasets.

      The analysis of the TOAs consists of two steps. In the first step {\sc
      Tempo2} is used to find the timing solution for a single pulsar. This
      includes possible jumps between datasets. Once the timing solution is
      obtained, the results are passed on to the Bayesian algorithm. The
      Bayesian algorithm then analytically marginalises all parameters of the
      timing model, including jumps, while using MCMC to explore the rest of the
      parameter space.

  \section{Results} \label{sec:results}
    Now that we have developed the necessary framework to analyse the TOAs, we
    apply the algorithm to the observations. In the following sub-sections we
    explain in detail how we selected the five pulsars that we already mentioned in
    Section \ref{sec:usedpulsars}, and we present the GWB upper limit
    we are able to calculate using observations of those pulsars.

    \subsection{Selecting the most constraining datasets} \label{sec:selection}
      For any pulsar, obtaining the timing solution and  timing residuals
      is the first step after obtaining the TOAs. The timing residuals
      of the pulsars used in this work are shown in Figure \ref{fig:allresiduals},
      and the parameters of the timing model are shown in the Appendix. The
      timing model also includes several jumps as some of these pulsars have been
      observed with several European telescopes.
      The timing solutions we find are quite consistent with the values already
      published in the literature. Given that we are solving for $56$
      parameters, it is to be expected that one or two parameters deviate at the
      $2$-$\sigma$ level. The only unexpected outlier we find is the proper
      motion in right ascension of J1713$+$0747, which deviates from
      \citet{Splaver2005} by over $5$-$\sigma$.
      Given that we are combining
      data of several telescopes, and that we do not take into account our red
      noise models in listing these timing solutions, we postpone exploring this
      difference to future work where the focus lies on
      investigating the statistics of the timing model parameters in the
      presence of red noise. Such an investigation is beyond the scope of this
      manuscript.

      With the model of the systematic contributions in place, we first perform
      the  analysis separately for each of the  datasets and obtain the posterior
      probability distribution for their intrinsic noise parameters, specified
      in Equation (\ref{eq:noisematrix}) of the previous section.  Note that at this stage of the
      analysis the contribution from a GWB is not yet included.  We determine a
      marginalised posterior for each pulsar
      as a function of the following parameter combinations:\newline
      1) EFAC vs. EQUAD\newline
      2) Red noise amplitude vs. red noise spectral index\newline
      In both cases, the posterior is marginalised over all parameters but two,
      and the resulting $2$-dimensional distribution is displayed as contours at
      the $1$-, $2$-, and $3$-$\sigma$ level (the regions where respectively
      $68\%$, $95\%$, and $99.7\%$ of the volume of the posterior is enclosed).
      
\begin{figure}
	\includegraphics[width=0.5\textwidth]{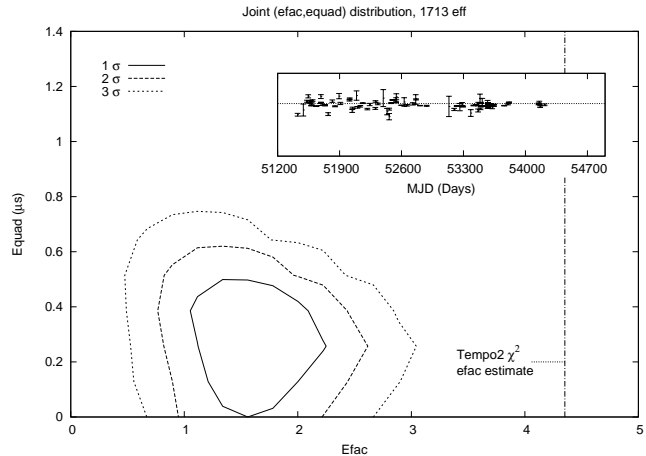}
	\caption{The marginalised posterior of J1713+0747 (Effelsberg), as a function of
	  the EFAC and EQUAD parameters.
	  The contours are at the $1$, $2$, and
	  $3$-$\sigma$ level, indicating a respective volume inside that region
	  of $68\%$, $95\%$, and $99.7\%$.}
	\label{fig:1713effefacequad}
      \end{figure}

\begin{figure}
	\includegraphics[width=0.5\textwidth]{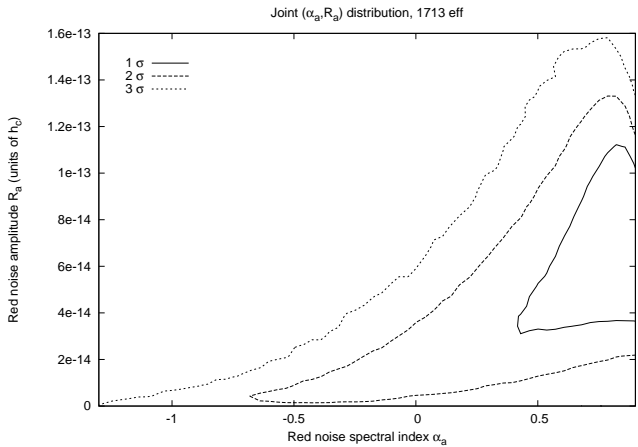}
	\caption{The marginalised posterior of J1713+0747 (Effelsberg), as a function of
	  the power-law red noise parameters: the amplitude and the spectral
	  index. The contours are at the $1$, $2$, and
	  $3$-$\sigma$ level, indicating a respective volume inside that region
	  of $68\%$, $95\%$, and $99.7\%$.}
	\label{fig:1713effrednoise}
      \end{figure}

      As an example we consider the TOAs of pulsar J1713+0747, which consist of data
      taken with Effelsberg and Westerbork.
      Here we focus on the marginalised
      posterior distributions that represent information about the Effelsberg
      TOAs; these distributions and the residuals are shown in Figs
      \ref{fig:1713effefacequad} and \ref{fig:1713effrednoise}. A traditional
      non-Bayesian analysis of the Effelsberg TOAs consists of a fit to the
      timing model with {\sc Tempo2}, which yields the best-fit parameters, the
      corresponding uncertainties, and a reduced $\chi^2$ statistic. The
      reduced $\chi^2$ is defined as:
      \begin{equation}
	\chi^2 = \frac{1}{n-m}\sum_{i=1}^{n}\frac{\left( t_i^{\text{obs}}
	- t_i^{fit}\right)^2}{\epsilon^2 \sigma_i^2},
	\label{eq:reducedchisqr}
      \end{equation}
      where $n$ is the number of observations, $m$ is the number of free
      parameters in the least-squares fit, $t_i^{\text{obs}}$ is the
      observed TOA, $t_i^{\text{fit}}$ is best-fit value of the TOA, $\sigma_i$
      is the TOA uncertainty of $t_i^{\text{obs}}$, and $\epsilon$ is the EFAC
      value. It is usual practice to set the EFAC such that the reduced
      $\chi^2=1$, which is accomplished by: $\epsilon =
      \sqrt{\chi^2(\epsilon=1)}$. For the J1713+0747 Effelsberg TOAs, we have
      $\chi^2(\epsilon=1) = 18.9$ and therefore $\epsilon = 4.35$.

      As can be seen
      from Figure \ref{fig:1713effrednoise}, a non-zero red noise component is
      required to describe the TOAs. The EQUAD parameter is consistent with
      $0$-amplitude according to Figure \ref{fig:1713effefacequad}, while the
      EFAC parameter is significantly lower than what a {\sc Tempo2} $\sqrt{\chi^2}$
      estimate would give.
      This tells us that no separate white-noise component is required to
      describe the TOAs: all the uncorrelated scatter can be assigned to the
      error bars on the TOAs. It is also of interest that in this case the
      EFAC parameter is much smaller, and indeed much closer to $1$, than
      the more traditional estimator $\sqrt{\chi^2}$. The data is
      reasonably well-modelled by just the presence of red noise.

      It is also worth noting that, due to practicalities having to do
      with hardware changes at the observatories, the TOAs of J1713+0747 end at
      an earlier epoch than the other 4 pulsars. Although in the future the
      inclusion of this data will obviously benefit the sensitivity to the GWB,
      we note that the GWB limit is not negatively effected by this lack of
      overlap of the TOAs between pulsars.

      The analysis of the TOAs of the other pulsars yields similar, but slightly
      different results. As can be seen in the appendix, some pulsars do have
      non-negligible white noise, and some do appear to have EFAC values
      significantly different from $1$. As of yet we do not have a complete
      explanation for the exact form of the marginalised posteriors.

      We present the marginalised posterior as a function of the red noise
      parameters in an intuitive way: as pointed out in Section
      \ref{sec:usedpulsars}
      we use the same units for the red noise
      amplitude and red noise spectral index as we use for the GWB parameters.
      For the analysis of TOAs of just one pulsar, the red noise can now be
      thought of as if it was generated solely by a GWB with a certain amplitude
      and spectral index. In this case, the marginalised posterior for the red
      noise parameters shows us what upper limit we are able to place on the GWB
      amplitude as a function of spectral index.

      We choose a $3$-$\sigma$ threshold of $R_{a} \leq 10^{-13}$ at a spectral index of
      $\alpha_{a}=-2/3$. Based on the marginalised posteriors of all the EPTA
      pulsars, we can decide whether a particular dataset can put a constraint
      on the GWB lower than this or not. Using this threshold we include five 
      pulsars in our final analysis. These five significantly outperform the
      other pulsars in terms of how well they can limit the GWB amplitude, and
      we do not expect to gain any significant sensitivity by
      including more pulsars in our current archival data sets. The residuals of the pulsars we use in our
      combined analysis are shown in Figure \ref{fig:allresiduals}. More
      datasets will be added after some extensive and detailed recalibration
      procedure of existing datasets.

    \subsection{GWB upper limit} \label{sec:upperlimit}
      Now that we have selected our pulsars that can significantly contribute to
      a GWB limit, we are in the position to infer the amplitude and spectral
      index of the GWB. Our model of the combined data of the five pulsars we
      selected in Section \ref{sec:selection} consists of all sources we
      included in the analysis for the single pulsars, and an extra source that
      corresponds to the GWB. As discussed in Section \ref{sec:bayesianmodels},
      the GWB source is a power-law correlated between pulsars as described by
      Equation (\ref{eq:zetaab}).

\begin{figure}
	\includegraphics[width=0.5\textwidth]{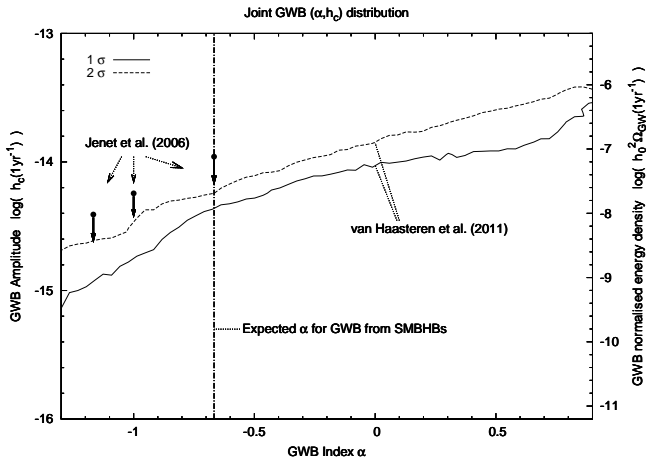}
	\caption{The marginalised posterior distribution as a function of the
	  GWB amplitude, and spectral index. The contours marked by 'van
	  Haasteren et al. (2011)' are the results of this work at the $1$-$\sigma$ and
	  $2$-$\sigma$ level, indicating a respective volume inside that region of
	  $68\%$, and $95\%$.  The vertical dash-dotted line at $\alpha=-2/3$
	  shows where we expect a GWB generated by supermassive black-hole
	  binaries. The most recent published limits are shown as the three
	  upper limit arrows pointing down, marked by 'Jenet et al. (2006)'.}
	\label{fig:gwblimit}
      \end{figure}

      As before, we use MCMC to sample the posterior distribution while
      analytically marginalising over the timing model;
      now the analytic 
      marginalisation happens
      simultaneously for the timing models of the five pulsars.
      In  Figure
      \ref{fig:gwblimit} we present the posterior, marginalised over all
      parameters except the GWB amplitude and 
      spectral index. In the same figure we also show the 
      PPTA published values of the GWB limit \citep{Jenet2006}. For the expected
      spectral index for a GWB generated by a large number of supermassive
      black-hole binaries, $\alpha=-2/3$, we find a 95\% confidence GWB upper limit of
      $h_c(1\text{yr}) \leq 6\times 10^{-15}$. This is smaller by a factor of 1.8 than the
      previously published PPTA limit.

\begin{figure}
	\includegraphics[width=0.5\textwidth]{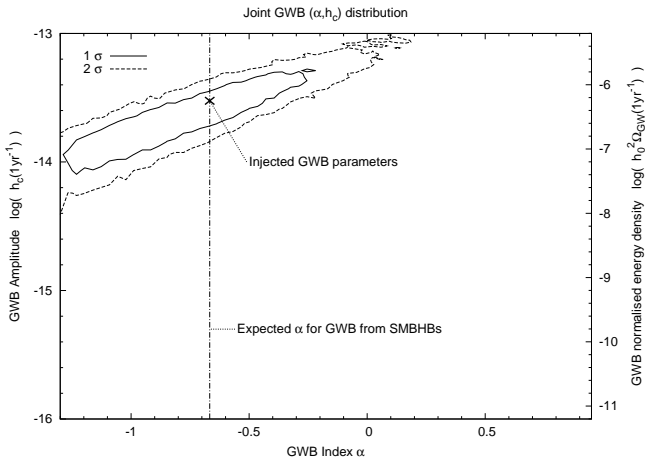}
	\caption{Same marginalised posterior distribution as in \ref{fig:gwblimit}, but here we have injected the
	  residuals of a simulated GWB with amplitude $h_c(1\text{yr}) = 30\times 10^{-15}$ in the
	  data.}
	\label{fig:gwblimitwithgw}
      \end{figure}

      As a cross-check with other codes, and to verify that we are definitely
      sensitive to the level of the limit we have calculated here, we perform an
      additional test. We use the {\sc Tempo2} plug-in GWbkgrd \citep{Hobbs2009}
      to generate simulated timing residuals as produced by a GWB with an
      amplitude of $h_c(1\text{yr})$. We then create a new set of TOAs, consisting of the
      values of the simulated timing residuals added to the values of the
      observed TOAs of the five pulsars that we have analysed in this section. We
      then redo
      the whole analysis. Current PTAs aim to reach sensitivities in the order
      of $h_c(1\text{yr}) = 10^{-15}$ in the future \citep{Jenet2005}, which is over five 
      times more sensitive than the limit we achieve here. In the case that the GWB
      just happens to be at the $2$-$\sigma$ level of our current limit, we
      demonstrate what such a fivefold increase in sensitivity could do for our
      ability to measure the GWB parameters by adding a signal of $h_c(1\text{yr}) =
      30\times 10^{-15}$ to our current TOAs. The result is shown in Figure
      \ref{fig:gwblimitwithgw}. We find that the results are consistent with the
      input parameters of the simulated GWB, and that we can reliably detect a
      GWB in this case\footnote{We note that, although such a detection is
      consistent with a GWB, we would need more pulsars to exclude the
      possibility that some other effect is causing the correlated signal we
      detect here.}. The values of the GWB parameters we have used to
      simulate the GWB lie within the $1$-$\sigma$ credible region of Figure
      \ref{fig:gwblimitwithgw}.
      
  \section{Implications} \label{sec:implications}
    The analysis performed in this work puts an upper limit on a GWB with a
    power-law characteristic strain spectrum $h_{c} =
    A(f/\hbox{yr}^{-1})^{\alpha}$. In the literature, upper limits are typically
    quoted for various values of $\alpha$, where the considered $\alpha$ depends
    on the physics responsible for generation of the GWB. A useful feature of
    our approach is that we are able to measure $\alpha$ for a strong enough
    GWB (see vHLML for a discussion). The extra degree of freedom in our model,
    $\alpha$, necessarily changes the interpretation of the posterior to some
    extent. We interpret the $2$-$\sigma$ contour in our plot of the marginalised
    posterior as the upper limit on the GWB as a function of $\alpha$. Fixing
    $\alpha$ and re-evaluating the $2$-$\sigma$ limit based on the posterior for
    $A$ only does not significantly alter our results.

    In this section, we briefly discuss the implications of the new
    upper limits in the context of two different mechanisms for generation
    of the GWB: binaries of supermassive black holes, and cosmic
    strings.\newline

    \subsection{Supermassive black hole binaries}
      Several authors discuss the characteristic strain spectrum generated by an
      ensemble of supermassive black holes (SMBHBs) distributed throughout the Universe
      \citep{Begelman1980, Phinney2001, Jaffe2003, Wyithe2003}. They show that
      the characteristic strain spectrum generated by such black hole binaries
      can well be approximated by a power-law:
      \begin{eqnarray} \label{eq:charstrain2}
	h_c&=&h_{\text{1yr}}\left(f\over \hbox{yr}^{-1}\right)^{-2/3},
      \end{eqnarray}
      where $h_{\text{1yr}}$ is a model-dependent constant. Though the 
      form of the characteristic strain, the power-law, is quite general among
      the different SMBHB assembly models
      the authors use in their work, the parameterisations
      and assumptions about other physical quantities differ between all
      investigators. The predicted $h_{\text{1yr}}$ therefore differs depending
      on what SMBHB assembly scenarios were assumed.

      Recently, \citet{Sesana2008} have extensively investigated a wide variety of
      assembly scenarios, including those considered in \citet{Jenet2006}. For our
      purposes in this work, their most important result is an estimate of
      $h_{\text{1yr}}$ for all models\footnote{The model for the GWB that \citet{Sesana2008} use
      is a broken power-law. Their $h_{\text{1yr}}$ therefore has a
      slightly different meaning, and our quoted value should be taken as a
      crude approximation.}. In calculating this value, they take into
      account the uncertainties of the key model parameters for different
      scenarios, and come up with
      $h_{\text{1yr}} \approx 2 \times 10^{-16} - 4\times 10^{-15}$. We are less
      than a factor of two away from this range, so we foresee that we can start
      to rule out some models in the near future.

      Two more results of \citet{Sesana2008} are interesting with respect to the
      limit presented in this work. The first is that the frequency dependence
      of the GWB is expected to be steeper than a power-law $\propto f^{-2/3}$
      for frequencies $f \gtrsim 10^{-8}$ Hz. The steepness depends on the
      chosen model. We have incorporated a varying spectral index $\alpha$ in
      our current analysis, and since we are not yet able to detect the GWB, we
      postpone a more thorough investigation of the exact dependence of $h_c$ on
      $f$ to later work with even better datasets.  The second interesting
      result is that in the frequency range of $10^{-8}$ Hz $\leq f \leq
      10^{-7}$ Hz, the GWB might be dominated by single sources. In that case, a
      search for just a certain characteristic strain spectrum is not
      appropriate, and we note that further investigation is required in this
      regard.

    \subsection{Cosmic strings}
      Several authors have suggested that oscillating cosmic string loops will
      produce gravitational waves \citep{Vilenkin1981, Damour2005, Olmez2010}.
      \citet{Damour2005} have used a semi-analytical
      approach to derive the characteristic strain $h_c$ of the GWB generated by
      cosmic strings:
      \begin{eqnarray}
	h_{c}(f) &=& 1.6\times 10^{-14}c^{1/2}p^{-1/2}\epsilon_{\text{eff}}^{-1/6}
	\nonumber \\
	& &\times (h/0.65)^{7/6}\left( \frac{G\mu}{10^{-6}}\right)^{1/3}\left(
	\frac{f}{\text{yr}^{-1}} \right)^{-7/6},
	\label{eq:stringgwb}
      \end{eqnarray}
      where $\mu$ is the string tension, $G$ is Newton's constant, $c$ is the
      average number of cusps per loop oscillation, $p$ is the reconnection
      probability, $\epsilon_{\text{eff}}$ is the loop length scale factor, and
      $h$ is the Hubble constant in units of $100 \text{km
      s}^{-1}\text{Mpc}^{-1}$. Usually, the dimensionless combination $G\mu$ is
      used to characterise the string tension. Theoretical predictions of string
      tensions are $10^{-11} \leq G\mu \leq 10^{-6}$ \citep{Damour2005}.

      From the above expression for the characteristic strain generated by
      cosmic strings, we see that this is again a power-law, but now with
      $\alpha=-7/6$. Using a standard model assumption that $c=1$, the facts
      that $p$ and $\epsilon_{\text{eff}}$ are less than one, and that $h$ is
      expected to be greater than $0.65$, we can safely use our derived upper
      limit on $h_c$ for $\alpha=-7/6$ to limit the string tension: $G\mu \leq
      4.0 \times 10^{-9}$. This already places interesting constraints on the
      theoretical models, and in a few years the EPTA will be able to place much
      tighter restrictions in the case of a non-detection of a GWB: with only a
      factor of five decrease of the upper limit, we would be able to completely
      exclude the $10^{-11} \leq G\mu \leq 10^{-6}$ range mentioned in
      \citet{Damour2005}.

  \section{Conclusion and discussion}
    In this paper we have developed the methodology on how to handle combined PTA
    datasets of several telescopes and how to robustly calculate a corresponding
    upper limit on the GWB. Our Bayesian approach has handled in a
    straightforward way different data sets of
    varying duration, regularity, and quality.
    The current upper limit on the
    GWB,
    calculated with EPTA data, is $h_{c} \leq 6\times 10^{-15}$ in the case of
    $\alpha = -2/3$, as predicted for a GWB created by an ensemble of
    supermassive BH binaries.
    More generally, the analysis has resulted in a marginalised
    posterior as a function of the parameters of the GWB: the GWB amplitude and
    the spectral index.

    Due to hardware and software upgrades at the EPTA observatories, and due to
    the ever increasing time baseline of the data, we expect the sensitivity to
    increase greatly over the next few years. Especially the combination of the
    EPTA data sets with the data of the other PTA projects seems promising.

    The raw telescope data must first undergo careful reduction and modelling
    before it can be analysed by the Bayesian algorithm.
    We have provided some
    discussion of these processes and have motivated our choice of model
    for the TOAs. As part of our analysis, we have studied  the probability
    distribution of the pulsar noise parameters, and highlighted the crucial
    importance of precise characterisation of the red component of pulsar timing
    noise.

  \section*{Acknowledgements}
    This research is supported by the Netherlands organisation for Scientific
    Research (NWO) through VIDI grant 639.042.607.

    We are very grateful to all staff at the Effelsberg, Westerbork, Jodrell
    Bank and Nan\c cay radio telescopes for their help with the observations
    used in this work. Part of this work is based on observations with the
    $100$-m telescope of the Max-Planck-Institut f\"ur Radioastronomie (MPIfR)
    at Effelsberg. Access to the Lovell telescope is supported through an STFC
    rolling grant. The Nan\c cay radio telescope is part of the Paris
    Observatory, associated with the Centre National de la Recherche
    Scientifique (CNRS), and partially supported by the R\'egion Centre in
    France. The Westerbork Synthesis Radio Telescope is operated by the
    Netherlands Foundation for Research in Astronomy (ASTRON) with support from
    the NWO.

    We would like to thank our referee, R.N. Manchester, for his careful review
    of this manuscript, and for his useful comments.


  \bibliographystyle{mn2e.bst}


  \section*{Appendix A}
    Here we show the timing solutions of all datasets used in this work,
    combined with the posterior distributions for the timing noise.


\begin{table*}
  \footnotesize
  \centering
  \begin{minipage}[t]{\textwidth}
  \label{tab:solutions}
  \begin{tabular}{
        l@{\hspace{0mm}}
        c@{\hspace{1mm}}
        c@{\hspace{1mm}}
        c@{\hspace{1mm}}
        c@{\hspace{1mm}}
        c@{\hspace{1mm}}
    }
    \hline\hline\\*[-2ex]
      Pulsar name\dotfill & J0613$-$0200 & J1012$+$5307 & J1713$+$0747  \\
    \hline\\*[-2ex]
    \multicolumn{2}{c}{Fit and data set} & &  \\
    \hline\\*[-2ex]
      Telescopes used \dotfill & NRT & NRT & EFF \& WSRT \\
      MJD range \dotfill & 53367 - 55012 & 53443 - 55030 & 51426 - 54637 \\
      Number of TOAs \dotfill & 280 & 107 & 195 \\
      Rms timing residual (ns) \dotfill & 912 & 769 & 396 \\
      Reduced $\chi^2$ value \dotfill & 1.00 & 1.00 & 1.13 \\
      Epoch \dotfill & 54189 & 54236 & 53031 \\
      \hline\\*[-2ex]
      \multicolumn{2}{c}{Measured Quantities} && \\
      \hline\\*[-2ex]
      Right ascension, $\alpha$ (J2000)\dotfill &  06:13:43.97385(4) & 10:12:33.43241(10) & 17:13:49.530782(3) \\
      Declination, $\delta$ (J2000)\dotfill & $-$02:00:47.0720(12) & +53:07:02.665(2) & +07:47:37.52343(8) \\
      Pulse freq., $\nu$ (s$^{-1}$)\dotfill & 326.600562095168(13) & 190.26783448248(14) & 218.811840486637(30) \\
      Derivative of pulse freq., $\dot{\nu}$ (s$^{-2}$)\dotfill & $-$1.02281(3)$\times 10^{-15}$ & $-$6.1998(4)$\times 10^{-16}$ & $-$4.0836(2)$\times 10^{-16}$ \\
      PM in RA, $\mu_{\alpha}$ (mas\,yr$^{-1}$)\dotfill & 1.90(4) & 3.17(7) & 5.017(12) \\
      PM in DEC, $\mu_{\delta}$ (mas\,yr$^{-1}$)\dotfill & $-$10.31(9) & $-$24.96(9) & $-$3.96(3) \\
      Parallax, $\pi$ (mas)\dotfill & --- & --- & 0.915(7) \\
      Dispersion measure, DM (cm$^{-3}$pc)\dotfill & 38.77700 & 9.0176 & 15.9907 \\
      \\*[-2ex]
      Binary model\dotfill & DD & ELL1 & DD \\
      Orbital period, $P_b$ (d)\dotfill & 1.19851257534(5) & 0.60462272322(4) & 67.8253309255(20) \\
      Derivative of orbital period, $\dot{P_b}$\dotfill & --- & --- & --- \\ 
      Epoch of periastron, $T_0$ (MJD)\dotfill & 54189.019(6) & --- & 53014.9592(7) \\
      Projected sm. axis of orbit, $x$ (lt-s)\dotfill & 1.09144417(8) & 0.58181742(13) & 32.34242015(7) \\
      Longitude of periastron, $\omega_0$ (deg)\dotfill & 47.1(1.6) & --- & 176.2109(12) \\
      Orbital eccentricity, $e$\dotfill & 5.47(15) $\times 10^{-6}$ & --- & 7.49312(13) $\times 10^{-5} $ \\
      Time of ascending node (MJD)\dotfill & --- & 54236.2078302(3) & --- \\
      EPS1 ($\epsilon_1$), $e \sin \omega$\dotfill & --- & 1.18(5)$\times 10^{-5}$ & --- \\
      EPS2 ($\epsilon_2$), $e \cos \omega$\dotfill & --- & 2.20(5)$\times 10^{-5}$ & --- \\
      Sine of inclination angle, $\sin{i}$\dotfill & --- & --- & --- \\
      Companion mass, $M_c$ ($M_\odot$)\dotfill & --- & --- & --- \\ 
      \hline\\*[-2ex]
      \multicolumn{2}{c}{Assumptions} &&\\
      \hline\\*[-2ex]
      Clock correction procedure\dotfill & \multicolumn{3}{c}{TT(TAI)}\\
      Solar system ephemeris model\dotfill & \multicolumn{3}{c}{DE405}\\
    \hline\\*[-2ex]
  \end{tabular}
  \caption{
    The timing solutions for the pulsars used in this paper before
    applying the Bayesian algorithm. These solutions are determined using
    \textsc{Tempo2}, which uses the International Celestial Reference System and
    Barycentric Coordinate Time. As a result this timing model must be modified
    before being used with an observing system that inputs \textsc{Tempo} format
    parameters. See \citet{Hobbs2006} for more information.
    Note that the figures in parentheses are the nominal
    $1$-$\sigma$ \textsc{Tempo2} uncertainties, with EFACs included, and
    therefore do not include the red noise model. In the GWB limit calculation
    these respective parameters are marginalised over. Also, the dispersion
    measure quoted here results from combining these observations with EPTA data
    of other frequencies. These DM values are used in the dedispersion, but we
    didn't include all observations in our GWB analysis. We therefore have not
    fit for the DM here, and an error estimate cannot be given.}
  \end{minipage}
\end{table*}

\begin{table*}
  \footnotesize
  \centering
  \begin{minipage}[t]{\textwidth}
  \label{tab:solutions2}
  \begin{tabular}{
        l@{\hspace{0mm}}
        c@{\hspace{1mm}}
        c@{\hspace{1mm}}
        c@{\hspace{1mm}}
        c@{\hspace{1mm}}
        c@{\hspace{1mm}}
    }
    \hline\hline\\*[-2ex]
      Pulsar name\dotfill & J1744$-$1134 & J1909$-$3744 \\
    \hline\\*[-2ex]
    \multicolumn{2}{c}{Fit and data set} &  \\
    \hline\\*[-2ex]
      Telescopes used \dotfill & EFF \& NRT & NRT \\
      MJD range \dotfill & 51239 - 55001 & 53366 - 55127 \\
      Number of TOAs \dotfill & 159 & 113 \\
      Rms timing residual (ns) \dotfill & 444 & 134 \\
      Reduced $\chi^2$ value \dotfill & 1.05 & 1.00 \\
      Epoch \dotfill & 53120 & 54247 \\
      \hline\\*[-2ex]
      \multicolumn{2}{c}{Measured Quantities} & \\
      \hline\\*[-2ex]
      Right ascension, $\alpha$ (J2000)\dotfill &  17:44:29.391592(7) & 19:09:47.437982(5) \\
      Declination, $\delta$ (J2000)\dotfill & $-$11:34:54.5762(6) & $-$37:44:14.3176(2) \\
      Pulse freq., $\nu$ (s$^{-1}$)\dotfill & 245.426119777227(4) & 339.31568732355(1) \\
      Derivative of pulse freq., $\dot{\nu}$ (s$^{-2}$)\dotfill & $-$5.3817(4)$\times 10^{-16}$ & $-$1.614853(8)$\times 10^{-15}$ \\
      PM in RA, $\mu_{\alpha}$ (mas\,yr$^{-1}$)\dotfill & 18.817(10) & $-$9.490(11) \\
      PM in DEC, $\mu_{\delta}$ (mas\,yr$^{-1}$)\dotfill & $-$9.30(6) & $-$35.89(4) \\
      Parallax, $\pi$ (mas)\dotfill & 2.602(10) & 1.01(7) \\
      Dispersion measure, DM (cm$^{-3}$pc)\dotfill & 3.13632 & 10.37877 \\
      \\*[-2ex]
      Binary model\dotfill & --- &  ELL1 \\
      Orbital period, $P_b$ (d)\dotfill & --- & 1.53349947490(6) \\
      Derivative of orbital period, $\dot{P_b}$\dotfill & --- & 3.5(5)$\times 10^{-13}$ \\ 
      Epoch of periastron, $T_0$ (MJD)\dotfill & --- & --- \\
      Projected sm. axis of orbit, $x$ (lt-s)\dotfill & --- & 1.89799108(11) \\
      Longitude of periastron, $\omega_0$ (deg)\dotfill & --- & --- \\
      Orbital eccentricity, $e$\dotfill & --- & --- \\
      Time of ascending node (MJD)\dotfill & --- &  54247.169903748(15) \\
      EPS1 ($\epsilon_1$), $e \sin \omega$\dotfill & --- &  6.4(5.5)$\times 10^{-8}$\\
      EPS2 ($\epsilon_2$), $e \cos \omega$\dotfill & --- & $-$3(3)$\times 10^{-8}$ \\
      Sine of inclination angle, $\sin{i}$\dotfill & --- & 0.9980(3) \\
      Companion mass, $M_c$ ($M_\odot$)\dotfill & --- & 0.208(7) \\ 
      \hline\\*[-2ex]
      \multicolumn{2}{c}{Assumptions} &\\
      \hline\\*[-2ex]
      Clock correction procedure\dotfill & \multicolumn{2}{c}{TT(TAI)}\\
      Solar system ephemeris model\dotfill & \multicolumn{2}{c}{DE405}\\
    \hline\\*[-2ex]
  \end{tabular}
  \caption{
  Same as table 2.}
  \end{minipage}
\end{table*}

  \begin{figure*}
    \includegraphics[width=0.9\textwidth]{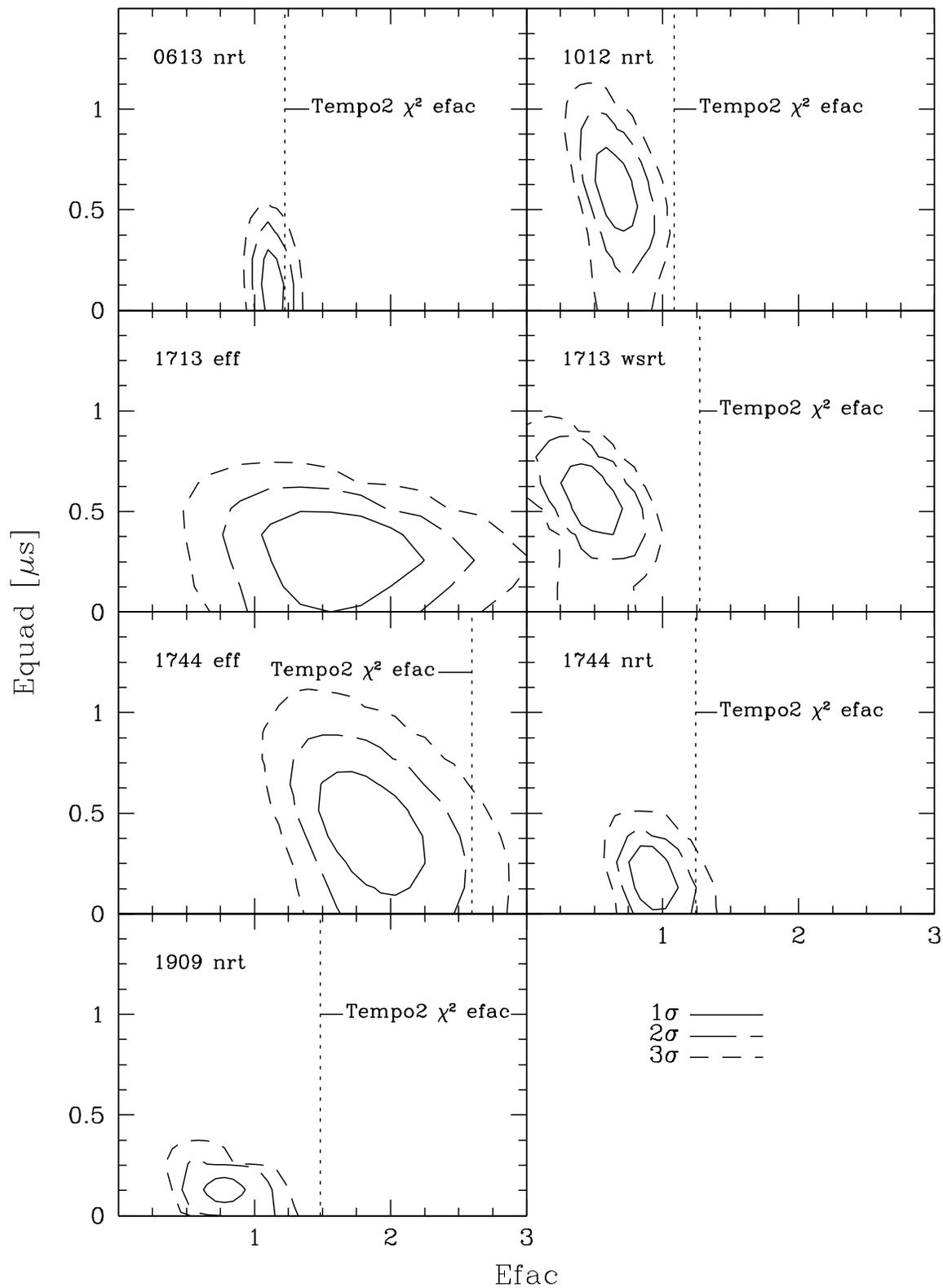}
    \caption{The marginalised posteriors of all datasets, as a function of
      the EFAC and EQUAD parameters.
      The contours are at the $1$, $2$, and
      $3$-$\sigma$ level, indicating a respective volume inside that region
      of $68\%$, $95\%$, and $99.7\%$. For the J1713-0747 posterior, the {\sc
      Tempo2} $\chi^2$ estimate is not shown because it has the off-scale value
      of $4.4$.}
    \label{fig:1909ncyefacequadapp}
  \end{figure*}

  \begin{figure*}
    \includegraphics[width=0.9\textwidth]{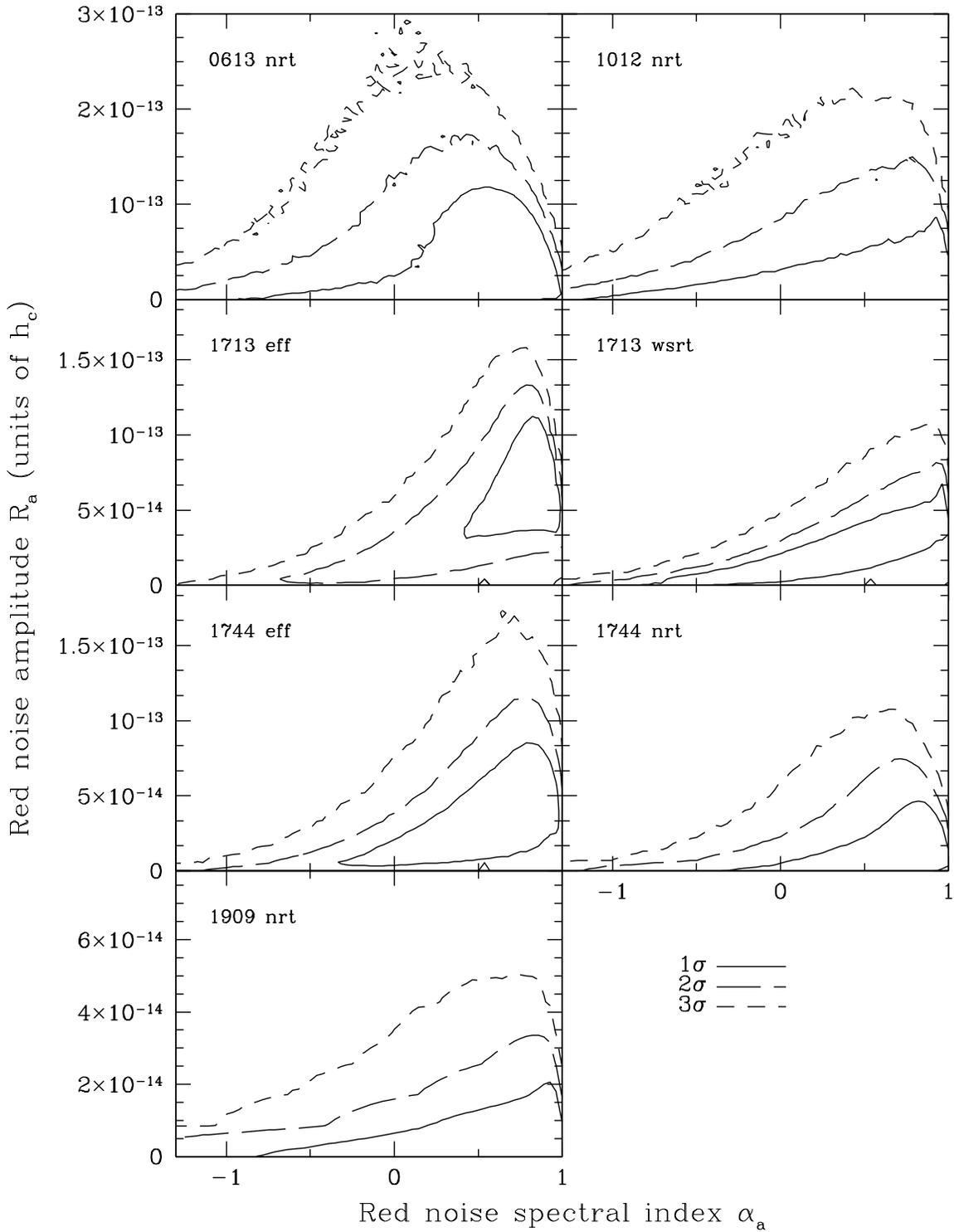}
    \caption{The marginalised posterior of all datasets, as a function of
      the power-law red noise parameters: the amplitude and the spectral
      index. The contours are at the $1$, $2$, and
      $3$-$\sigma$ level, indicating a respective volume inside that region
      of $68\%$, $95\%$, and $99.7\%$. The more negative the value of $\alpha$,
      the steeper the power-law spectrum, with the spectrum approaching a white
      spectrum at the right of the plot. We also note that the amplitude of the
      red noise cannot be trivially scaled linearly to an rms value of the
      timing residuals.}
    \label{fig:1909ncyrednoiseapp}
  \end{figure*}


\end{document}